# Optical Field Recovery in Jones Space

Qi Wu, Yixiao Zhu, Hexun Jiang, Qunbi Zhuge, *Senior Member, IEEE*, and Weisheng Hu

*Abstract*—Optical full-field recovery makes it possible to compensate for fiber impairments such as chromatic dispersion and polarization mode dispersion (PMD) in the digital signal processing. For cost-sensitive short-reach optical networks, some advanced single-polarization (SP) optical field recovery schemes are recently proposed to avoid chromatic dispersion-induced power fading effect, and improve the spectral efficiency for larger potential capacity. Polarization division multiplexing (PDM) can further double both the spectral efficiency and the system capacity of these SP carrier-assisted direct detection (DD) schemes. However, the so-called polarization fading phenomenon induced by random polarization rotation is a fundamental obstacle which prevents SP carrier-assisted DD systems from polarization diversity. In this paper, we propose a receiver of Jones-space field recovery (JSFR) to realize polarization diversity with SP carrier-assisted DD schemes in Jones space. Different receiver structures and simplified recovery procedures for JSFR are explored theoretically. The proposed JSFR pushes the SP DD schemes towards PDM without extra optical signal-to-noise ratio (OSNR) penalty. In addition, the JSFR shows good tolerance to PMD since the optical field recovery is conducted before polarization recovery. In the concept-of-proof experiment, we demonstrate 448-Gb/s reception over 80-km single-mode fiber using the proposed JSFR based on 2×2 couplers. Furthermore, we qualitatively compare the optical field recovery in Jones space and Stokes space from the perspective of the modulation dimension.

## I. Introduction

COHERENT detection has been widely deployed in the long-haul transmission system during the last decade since it can support 4-dimentinal (4-D) modulation including both phase and polarization diversity [1]. At present, the advent of 6G era and metaverse puts forward higher requirements on capacity upgrade in short-reach optical networks, such as data center interconnection within distance of hundreds of kilometers. Compared with coherent detection, direct detection (DD) is preferred for cost-sensitive short-reach optical networks since it relaxes the requirement for narrow-linewidth and high-stable lasers [2]. However, the chromatic dispersion-induced power fading effect [3] prevents the traditional intensity modulation and direct detection (IM-DD) systems from the optical interconnects beyond 80-km. As a result, self-coherent DD schemes such as Kramers-Kronig receiver (KKR) [4], carrier-assisted differential detection (CADD) [5], asymmetric self-coherent detection (ASCD) [6], deep-learning enabled direct detection (DLEDD) [7], and carrier-assisted phase retrieval schemes (CAPR) [8] have been recently proposed to recover the single-polarization (SP) optical field with both intensity and phase. These schemes are thus capable of receiver-side digital channel impairment compensation such as chromatic dispersion (CD). Polarization diversity should be the next goal that is required to be fulfilled since it doubles the system capacity and spectral efficiency.

For polarization division multiplexing (PDM), the polarization fading phenomenon induced by random polarization rotation is a fundamental obstacle [2, 9-11] for these carrier-assisted DD schemes. The polarization fading refers to that the optical carrier cannot be split equally into X- or Y-polarization using a simple polarization beam splitter (PBS) because the state of polarization (SOP) of the optical carrier after transmission is no longer controlled, which is different from the CD-induced power fading mentioned above. The SP optical field recovery is no longer fulfilled without a sufficiently strong carrier since these self-coherent detection schemes rely on the reference optical carrier to reconstruct the optical field digitally. One straightforward approach to avoid polarization fading is active polarization control [12]. Nevertheless, complicated device design and algorithms to adjust SOP are required for active polarization control, which increases the hardware cost and complexity for the receiver. In [13], an additional balanced-photodetector (BPD) and $3 \times 2$ multi-input-multi-output (MIMO) equalization are employed to make the system performance a weak function of the carrier SOP. The PDM single sideband (SSB) scheme with frequency-orthogonal carriers is proposed [11], which is based on optical bandpass filter (OBPF) or inter-leaver with sharp edges to separate the signals and carriers. Besides, Stokes-vector receiver (SVR) has been proposed in recent years to realize Stokes-space field recovery (SSFR) [14-22]. SSFR realizes 2-dimensional [15-18], 3-dimensional [19-21] and 4-dimensional [22] polarization modulation, and achieves high electrical and optical spectral efficiency. The SSFR conducts polarization de-rotation in Stokes-space before field recovery, and thus needs complex hardware structures.

In this work, we propose a dual-polarization (DP) field

Manuscript received xx, xx; revised xx, xx; accepted xx, xx. Date of publication xx, xx; date of current version xx, xx. This work was supported by National Natural Science Foundation of China under Grant 62001287 and by National Key R&D Program of China under Grant 2019YFB1803803. (Corresponding author: Yixiao Zhu)

Qi Wu, Yixiao Zhu, Hexun Jiang, Qunbi Zhuge and Weisheng Hu are with the State Key Laboratory of Advanced Optical Communication Systems and Networks, Department of Electronic Engineering, Shanghai Jiao Tong University, Shanghai, 200240, China (e-mail: sjtuwuqi@sjtu.edu.cn, yixiaozhu@sjtu.edu.cn, jianghexun@sjtu.edu.cn, qunbi.zhuge@sjtu.edu.cn, wshu@sjtu.edu.cn).

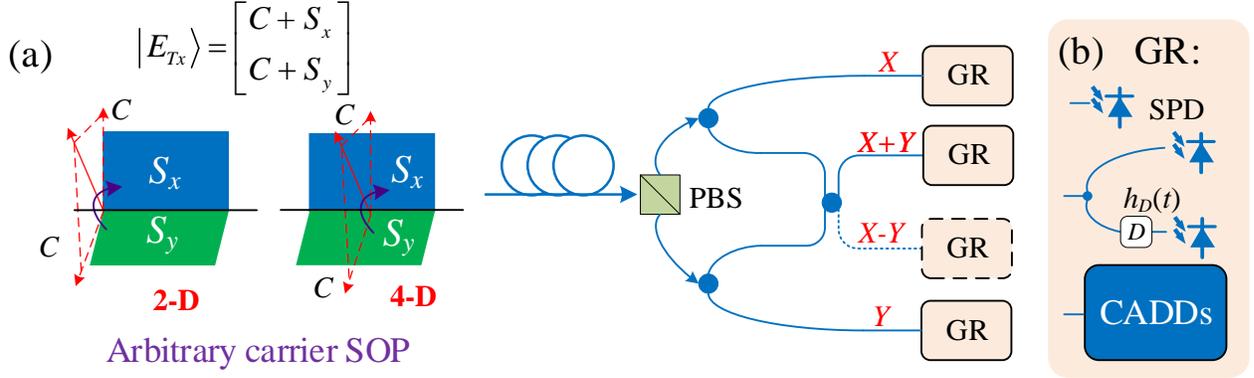

Fig. 1. (a) Receiver structures of the JSFR based on 2×2 couplers. (b) GR: generalized receiver. CADDs: CADD receiver and its simplified structures.

recovery scheme to fulfill polarization-fading-free detection in Jones space instead of Stokes space. The proposed Jones-space field recovery (JSFR) conducts field recovery first and then achieves polarization recovery using MIMO equalization, like coherent detection. JSFR can be divided into 2-dimensional (2-D) and 4-D JSFR after integrating with different SP self-coherent detection schemes. Specifically, KKR can be used for 2-D JSFR, and the advanced carrier-assisted single-polarization DD schemes including CADD, ASCD, DLEDD and CAPR can be implemented as generalized receivers to further achieve 4-D JSFR. The JSFR based on different structures is proposed and analyzed to make a parallel comparison between SSFR and JSFR. Besides, the simplification schemes of JSFR are also proposed to decrease the required hardware complexity. In the concept-of-proof experiment, we implement JSFR using single-ended photodetectors (SPDs) and KKR-based digital optical field recovery. 56 GBd DP-16-ary quadrature-amplitude-modulation (QAM) signal is successfully recovered after 80-km single-mode fiber (SMF) transmission.

The paper is organized as follows: Section II introduces the principle and simplification schemes of JSFR. Simulations are conducted in Section III to verify the principle of JSFR and investigate the PMD impact on JSFR. The experimental results of 448 Gb/s transmission over 80-km SMF are presented in Section IV to validate the effectiveness of 2-D JSFR. Finally, conclusions are drawn in Section V.

## II. PRINCIPLE OF JSFR

There have been many single-polarization carrier-assisted direct detection schemes proposed recently to avoid the power fading effect and improve spectral efficiency [4-8]. The receiver structures of these schemes are considered as the generalized receiver (GR) in the paper, shown in the Fig. 1(b). It can be implemented as one SPD for SSB signal [4], the two-branch phase retrieval receiver [23] for complex double sideband detection including ASCD, DLEDD, and central carrier-assisted phase retrieval (CCA-PR) schemes [6-8], or CADD receiver [5] and its simplified variants [24]. These schemes realize SP optical field recovery based on different receiver structures and optical field reconstruction algorithms, in which the power of reference optical carrier is the most important. Therefore, the carrier-to-signal power ratio (CSPR) is a crucial parameter for them. In brief, the optical field can be reconstructed digitally if the carrier power is sufficient large. For example, it is about 6 dB for KKR to satisfy the minimum phase condition [4], and then the optical field can be reconstructed for the SSB signal. The CSPR is about 10 dB for ASCD to eliminate signal-to-signal beat interference (SSBI) and detect a complex-valued double sideband (DSB) signal [6]. More generally, we use $C_{req}$ to represent for the required CSPR for various GRs to perform optical field recovery for both SSB and complex-valued DSB signals, and further introduce the principle of JSFR.

### A. Principle of field recovery in Jones space

The receiver structure of JSFR based on 2×2 couplers is shown in Fig. 1(a). The GR represents the receiver to achieve single-polarization field recovery. In this work, the available dimension is defined as the electrical spectral efficiency with respect to IM-DD system. As a consequence, 2-D JSFR and 4-D JSFR refer to the JSFR used to achieve optical field recovery of PDM-SSB, and PDM-DSB signals, respectively.

For the DP optical field recovery, the Jones vector of the transmitted carrier-assisted signal can be denoted as:

$$|E_{Tx}\rangle = \begin{bmatrix} C + S_x \\ C + S_y \end{bmatrix} \quad (1)$$

where $S_x$ and $S_y$ are the independent complex signals with equal power modulated on the X- and Y- polarization, respectively. $C$ is the optical carrier located on both X- and Y-polarization, respectively. Here, to facilitate the explanation of the principle, we assume that the carrier is located at the 45° between X- and Y- polarization. Note that the carrier SOP can be extended to arbitrary case, and the derivation is provided in Section II B. The CSPR is defined as the total power ratio between the carriers ($P(C)$) and the PDM signals of two polarizations ($P(S_{x/y})$), shown in Eq. (2).

$$CSPR = \{2 \cdot P(C)\} / \{P(S_x) + P(S_y)\} = P(C) / P(S_x) \quad (2)$$

Generally, the Jones transformation matrix representing polarization effect can be modeled as a 2×2 frequency-independent unitary matrix [25] in the form as:

$$\begin{bmatrix} \cos\alpha e^{j\theta} & -\sin\alpha \\ \sin\alpha & \cos\alpha e^{-j\theta} \end{bmatrix} \quad (3)$$

where $\alpha$ denotes the random polar angle and $\theta$ denotes the phase difference between X- and Y-polarization. The relation between the received optical Jones vector $|E_{Rx}\rangle$ and transmitted Jones vector $|E_{Tx}\rangle$ can be expressed as:

$$|E_{Rx}\rangle = \begin{bmatrix} \cos\alpha e^{j\theta} & -\sin\alpha \\ \sin\alpha & \cos\alpha e^{-j\theta} \end{bmatrix} \begin{bmatrix} C+S_x \\ C+S_y \end{bmatrix}$$
$$= \begin{bmatrix} \{\cos\alpha e^{j\theta} - \sin\alpha\}C + \cos\alpha e^{j\theta}S_x - \sin\alpha S_y \\ \{\sin\alpha + \cos\alpha e^{-j\theta}\}C + \sin\alpha S_x + \cos\alpha e^{-j\theta}S_y \end{bmatrix} \triangleq \begin{bmatrix} X \\ Y \end{bmatrix} \quad (4)$$

where $X/Y$ is defined as the complex optical field of receiver-side X/Y-polarization in Jones-space. At the receiver, $X$ is coupled with $Y$ employing a 2×2 optical coupler. The generated $X+Y$ and $X-Y$ are given as:

$$X+Y = \{2\cos\alpha\cos\theta\}C + (\cos\alpha e^{j\theta} + \sin\alpha)S_x + (\cos\alpha e^{-j\theta} - \sin\alpha)S_y$$
$$X-Y = \{-2\sin\alpha + j2\cos\alpha\sin\theta\}C + (\cos\alpha e^{j\theta} - \sin\alpha)S_x$$
$$\quad -(\cos\alpha e^{-j\theta} + \sin\alpha)S_y \quad (5)$$

The CSPRs of the four optical fields ($X$, $Y$, $X+Y$, $X-Y$) can be calculated as:

$$CSPR_X(\alpha,\theta) = (1 - 2\cos\alpha\sin\alpha\cos\theta) \cdot C_{req}$$
$$CSPR_Y(\alpha,\theta) = (1 + 2\cos\alpha\sin\alpha\cos\theta) \cdot C_{req}$$
$$CSPR_{X+Y}(\alpha,\theta) = (2\cos^2\alpha\cos^2\theta) \cdot C_{req} \quad (6)$$
$$CSPR_{X-Y}(\alpha,\theta) = (2 - 2\cos^2\alpha\cos^2\theta) \cdot C_{req}$$

It can be found that $CSPR_X + CSPR_Y = CSPR_{X+Y} + CSPR_{X-Y} = 2 \cdot C_{req}$, indicating that there are always two CSPR values larger than $C_{req}$ regardless of polarization rotation. We take an arbitrary SOP for example. When $\alpha$ is 45° and $\theta$ is $\pi/3$, $CSPR_X$, $CSPR_Y$, $CSPR_{X+Y}$, and $CSPR_{X-Y}$ are $0.5 \cdot C_{req}$, $1.5 \cdot C_{req}$, $0.25 \cdot C_{req}$, and $1.75 \cdot C_{req}$, respectively. Thus, the optical fields ($Y$, $X-Y$) can be recovered using corresponding GR and optical reconstruction algorithms, and then the DP optical full-field can be recovered using MIMO equalization.

Note that the derivation process is independent with the frequency of $C$. Therefore, $C$ can be located either at the edge or center of the signal spectrum, corresponding to 2-D and 4-D JSFR respectively.

### B. Arbitrary carrier SOP

To illustrate that the arbitrary carrier SOP at the transmitter side imposes no penalty on the JSFR scheme, we assume that there is a rotation angle $\xi$ between the carriers and signals, given as:

$$|E_{Tx}\rangle = \begin{bmatrix} C + \cos\xi S_x - \sin\xi S_y \\ C + \sin\xi S_x + \cos\xi S_y \end{bmatrix} \triangleq \begin{bmatrix} C + G_x \\ C + G_y \end{bmatrix} \quad (7)$$

The compound signals ($\cos\xi S_x - \sin\xi S_y$, $\sin\xi S_x + \cos\xi S_y$) can be generalized as $G_x$ and $G_y$. It is easy to prove that the powers of $G_x$ and $S_x$ are equal, shown as:

$$P(G_x) = P(\cos\xi S_x - \sin\xi S_y)$$
$$= \cos^2\xi P(S_x) + \sin^2\xi P(S_y) = P(S_x) \quad (8)$$

Here the cross terms are cancelled because $S_x$ and $S_y$ are independent zero-mean variables. For the new Jones vector $|E_{Tx}\rangle$, the derivation process is the same as Section II A. Once optical field is recovered in Jones space, $S_{x/y}$ can be obtained from $G_{x/y}$ with 2×2 equalization. Therefore, for JSFR, the SOP of the optical carrier requires no alignment at the transmitter.

### C. JSFR based on optical hybrid or 3×3 coupler

Note that the above receiver structure of JSFR is only based on three 2×2 couplers. It is possible to implement it based on one optical hybrid or 3×3 coupler. We name them as Scheme 1, 2, and 3, respectively. The detailed schemes and derivation process are analyzed in APPENDIX A.

### D. Simplification for JSFR

Some simplification can be conducted to decrease the hardware complexity of JSFR. For the branch to receive $X-Y$, it is not necessary because the photocurrents detected from $X-Y$ can be reconstructed using the other three photocurrents and a simple analog circuit. We take the GR as SPD or two-branch phase retrieval receiver [23] for example to illustrate the principle, given as:

$$|(X-Y) \otimes h(t)|^2 = |X \otimes h(t) - Y \otimes h(t)|^2$$
$$= 2|X \otimes h(t)|^2 + 2|Y \otimes h(t)|^2 - |(X+Y) \otimes h(t)|^2 \quad (9)$$

Here $\otimes$ is the convolution operation. $h(t)$ is $\delta(t)$ for one SPD or the dispersion function of dispersive element $h_D(t)$ for the dispersive branch, respectively. Therefore, theoretically, only three GRs are required for dual-polarization full-field recovery, which reduces the hardware complexity and cost of transceivers.

Another simplification for JSFR by enhancing carrier power is introduced in APPENDIX B.

## III. SIMULATIONS

In this section, we conduct simulations to numerically verify the principle of JSFR and characterize the impacts of CSPR, OSNR sensitivity, and PMD on the proposed JSFR. To match the following experiment, we recover the PDM-SSB signal with 56 GBd DP-16-QAM modulation using the proposed 2-D JSFR in Fig. 1. The GR is implemented as one SPD, and the optical field on each polarization is reconstructed digitally using KKR [4]. The simulations are performed at back-to-back (BTB) case. The roll-off factor of root raised cosine (RRC) filter is 0.01. The OSNR is set as 30 dB, and no electrical noise is imposed. The KKR operates at 8 samples-per-symbol (SPS) to avoid spectral broadening.

### A. Required CSPR for JSFR

It is known that the $C_{req}$ for KKR is about 6 dB with 16-QAM format in the simulation [4]. So, we set the CSPR as 6 dB to simulate the BERs versus SOPs. Here we use conventional DP direct detection receiver which consists of PBS and two SPDs without polarization control to compare the impact of polarization states with our proposed JSFR. As shown in Fig. 2(a), it can be observed that the polarization fading imposes huge damage to the DP direct detection receiver at certain SOPs, whose bit-error-rate (BER) can be as worse as $\sim 3.2 \times 10^{-1}$. When $\theta$ is $\pi/2$ or $3\pi/2$, the optical carrier is circularly polarized and thus the polarization fading will not occur. Then, we simulate the BERs as a function of SOPs with the proposed JSFR, shown in Fig. 2(b). The BER performance only fluctuates from $1.4\times10^{-3}$ to $1.8\times10^{-3}$ since the DP optical field can be reconstructed at any SOP.

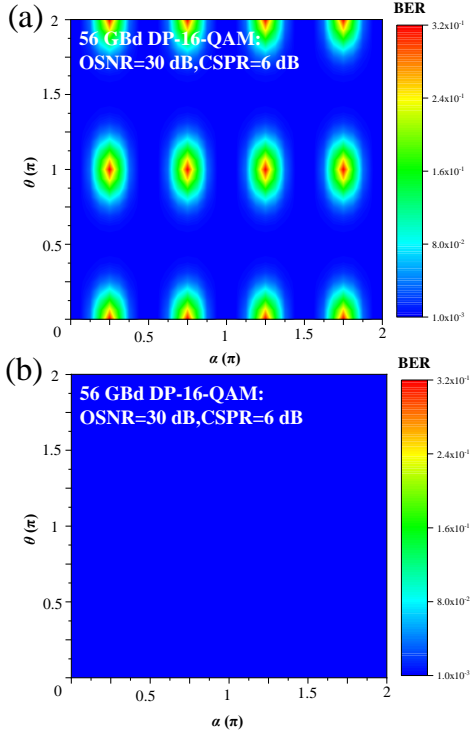

Fig. 2. (a) Simulated BERs versus SOPs using 2 SPDs only. (b) Simulated BERs versus SOPs using JSFR. The CSPR is 6 dB.

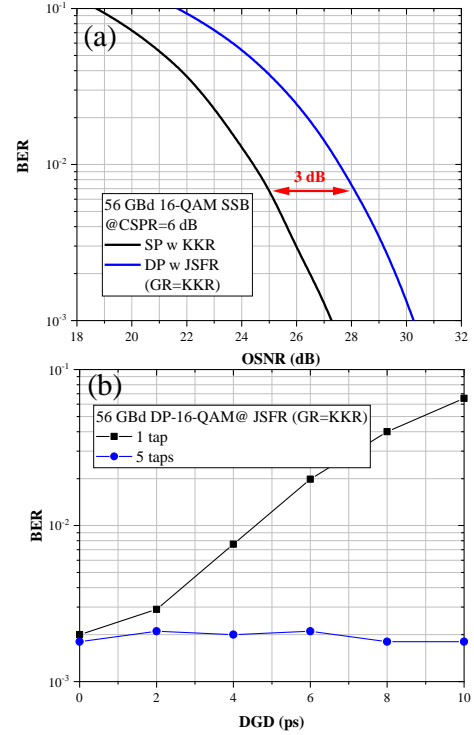

Fig. 3. (a) OSNR sensitivity between SP KKR and DP JSFR (GR=KKR). (b) Simulated BERs versus DGD with 1-tap and 5-tap equalizers.

### B. OSNR sensitivity between SP scheme and JSFR

We simulate the OSNR sensitivity of the SP KKR scheme and DP JSFR scheme, as shown in Fig. 3(a). The CSPR is 6 dB. It can be observed that the OSNR gap between SP and DP case is 3 dB, which fits well with the theoretical value. Therefore, the JSFR imposes no extra OSNR penalty on SP scheme while it realizes the polarization diversity for GR.

### C. PMD impact on JSFR

Different from the SSFR, the JSFR scheme first retrieves the phase of the optical field on the single-polarization, and then conducts polarization recovery through MIMO equalization. Fig. 3(b) shows the worst BERs as a function of differential group delay (DGD) between the principle state of polarizations using the 1-tap and 5-tap equalizers. The all-order PMD model is used here to emulate the DGD, and the fiber is divided into 15 sections to accurately model the PMD impact [26]. The PMD parameter is 0.1 ps/$\sqrt{\text{km}}$. Therefore, like coherent detection, the tolerance to PMD can be enhanced by simple linear equalization. The results imply that although the PMD imposes penalties on the GR in the SP case, like KKR [27], the penalty from PMD impact can be eliminated using JSFR.

## IV. EXPERIMENTS AND DISCUSSION

In this section, as a concept-of-proof experiment, we demonstrate and validate the principle of proposed 2-D JSFR using KKR-based SP field recovery, and GR is thus one SPD. The 4-D JSFR will be investigated in our future work.

### A. Experimental setup and DSP stacks

The experimental setup together with DSP stacks of 2-D JSFR is shown in Fig. 4. At the transmitter, after the generated bits are mapped to 16-QAM symbols, the 56 GBd DP-16-QAM sequences are shaped by a RRC filter with a roll-off factor of 0.01, leading to 56.56 GHz bandwidth. The I/Q components of the signal are generated by an arbitrary waveform generator (AWG, Keysight M8194A) with 45 GHz 3-dB bandwidth, operating at 120 GSa/s, and then loaded into an integrated DP-IQ modulator of 40 GHz 3dB bandwidth to modulate the light from external cavity laser (ECL1) (1549.98 nm). ECL2 (1550.21 nm) is used as the optical carrier, and the variable optical attenuator (VOA1) is set to adjust the CSPR. The polarization controller (PC1) after VOA1 is used to adjust the SOP of the optical carrier to 45° to match the transmitted Jones vector in Eq. (1). Note that JSFR is available for arbitrary carrier SOP. Before being launched into an 80-km SMF link, the combined SSB signal is amplified by an erbium-doped fiber amplifier (EDFA).

At the receiver, the signal is first amplified, and filtered using an OBPF to remove the out-of-band optical noise. The OSNR and SOPs of optical signal are adjusted for testing the system performance by VOA2 and PC2, respectively. After being split by a PBS, the generated optical fields $X$, $Y$, and their coupling term $X+Y$ and $X-Y$, are detected by 70 GHz bandwidth PDs (XPD3120R) and then captured by a real-time digital storage oscilloscope (DSO) with a 59 GHz brick-wall electrical bandwidth (Tektronix DPO75902SX) operating at 200 GSa/s for offline DSP. In this proof-of-concept experiment, we use one SPD to receive *X-Y* branch directly since there is a relative optical delay between the optical path (OP1) and OP2, which needs to be calibrated in advance like SSFR [22]. It is easy to keep OP1 and OP2 the same length with on-chip photonic integration technology [28].

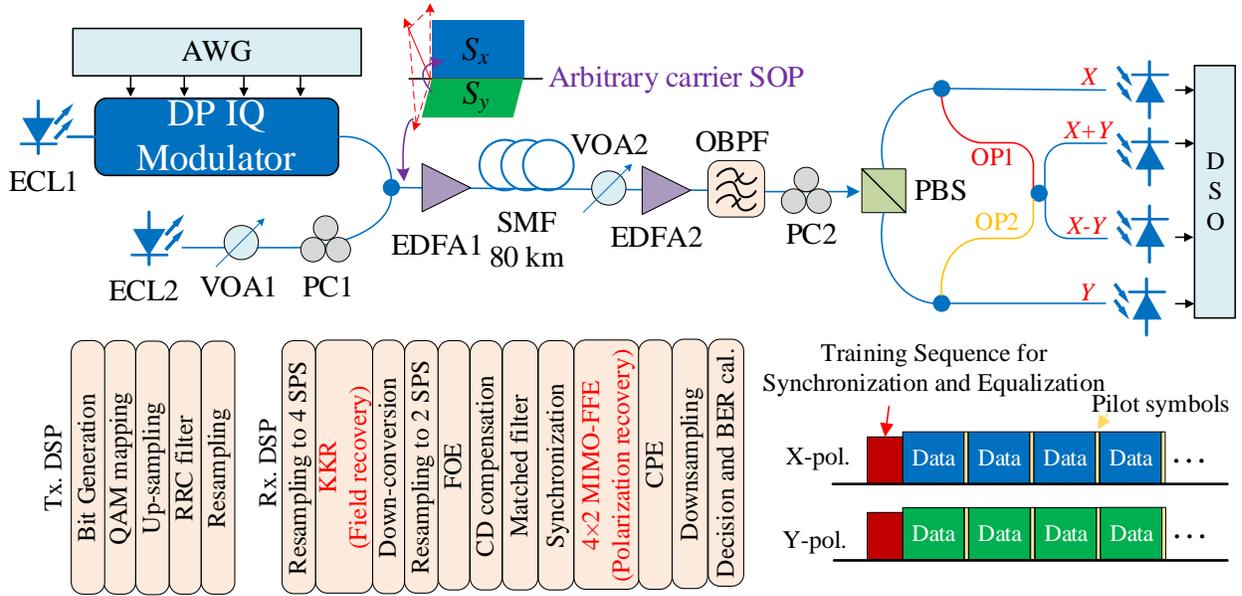

Fig. 4. Experimental setup and DSP stacks. ECL: external cavity laser; AWG: arbitrary waveform generator; EDFA: erbium-doped fiber amplifier; SMF: single-mode fiber; VOA: variable optical attenuator; OBPF: optical band-pass filter; PC: polarization controller; PD: photodetector. DSO: digital storage oscilloscope;

At the receiver DSP, the detected waveforms are firstly resampled to 4 SPS, and then we achieve optical field recovery using KKR. After KKR-based field recovery, the waveforms from 4 channels are down-converted to baseband and resampled to 2 SPS. Then, the residual frequency offset is estimated by the 4-th power method, and CD compensation is conducted [1]. After the matched filter, the synchronization is performed based on the training sequences located at the head of frame. These signals are fed into 4×2 time-domain MIMO equalizers for inter-symbol-interference cancellation and polarization demultiplexing simultaneously, like coherent detection. The MIMO equalization are based on complex-valued feed forward equalizers (FFE). The tap number is 51, and the equalizer taps are updated by using the recursive least square algorithm based on 512 training symbols. Then, the carrier phase estimation (CPE) based on pilot symbols and blind phase search [1] is conducted to eliminate phase noise. After down-sampling and decision, the BER is calculated. The frame structure is shown in Fig. 4. The symbol lengths of training sequences and payload data are 512 and 22400, respectively. The percentage of pilot symbols for CPE is 0.4%.

*B. Transmission performance*

First, we sweep the CSPR with 56 GBd SP-16-QAM SSB modulation and KKR-based optical field recovery to determine the $C_{req}$ for optical field reconstruction. The results are shown in Fig.5. The optimal CSPR is about 12 dB. Thus, we set the CSPR as 12 dB to ensure compliance with the principle of JSFR.

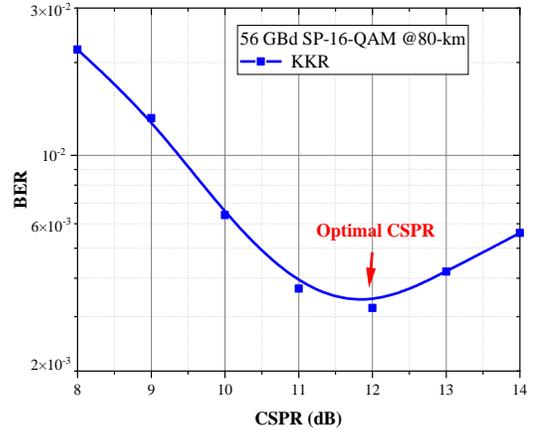

Fig. 5. Measured BERs versus CSPR with SSB modulation and KKR.

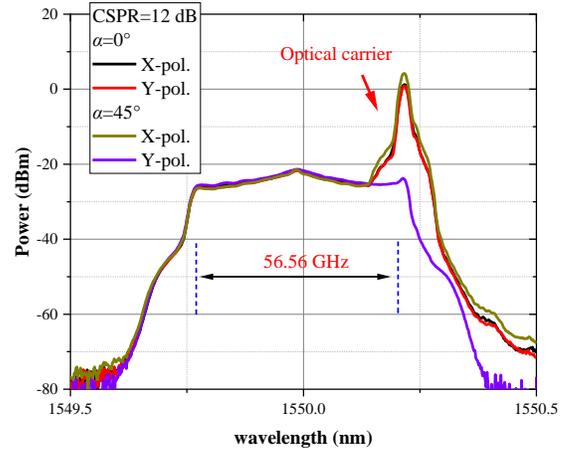

Fig. 6. Optical spectra of X- and Y-polarization signal with different polarization angles $\alpha$. The CSPR is 12 dB.

The signal is then switched to 56 GBd DP-16-QAM with PDM-SSB modulation and then we investigate the performance of JSFR with different polarization angles. The polarization

TABLE I
HARDWARE COMPLEXITY COMPARISON BETWEEN SSFR AND JSFR

| Scheme | 2-D modulation [18, 21] | 4-D modulation [22] |
|---|---|---|
| SSFR | Scheme **S-2D-1**: 2 couplers (2×2), OH, 3 BPDs | 2 OHs, 2 couplers (1×3), 4 BPDs, 2 SPDs |
| | Scheme **S-2D-2**: 2 couplers (2×2), OH, 4 SPDs | |
| | Scheme **S-2D-3**: 2 couplers (2×2, 3×3), 4 SPDs | |
| JSFR | Scheme **J-2D-1**: 3 couplers (2×2), 3 or 4 SPDs | Scheme **J-4D-1**: 3 couplers (2×2), 3 or 4 GRs |
| | Scheme **J-2D-2**: OH, 3 or 4 SPDs | Scheme **J-4D-2**: OH, 3 or 4 GRs |
| | Scheme **J-2D-3**: 3×3 coupler, 3 SPDs | Scheme **J-4D-3**: 3×3 coupler, 3 GRs |

OH: optical hybrid; 2×2: 2×2 coupler; 3×3: 3×3 coupler; SPD: single-ended photodetector; BPD: balanced photodetector; GR: generalized receiver.

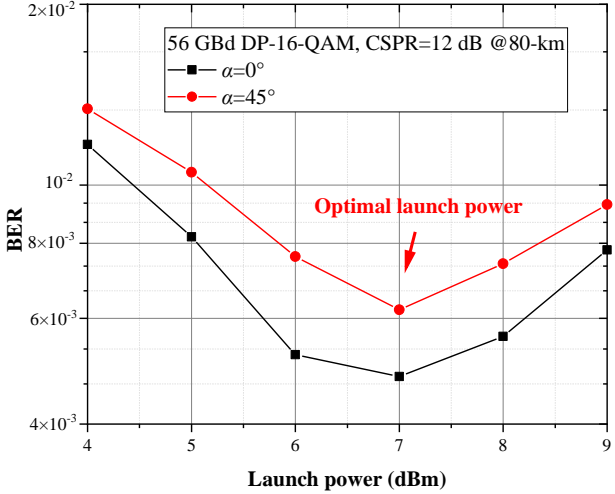

Fig. 7. Measured BERs versus launch power under 12 dB CSPR.

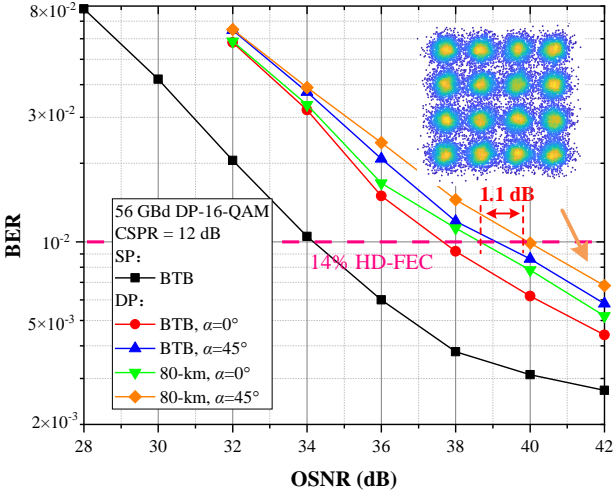

Fig. 8. Measured BERs versus OSNR with different polarization angles $\alpha$ at BTB and 80 km transmission case.

angle $\alpha$ is controlled by PC2. The received optical spectra after PBS with different $\alpha$ are shown in Fig. 6. $\alpha=0°$ and $\alpha=45°$ correspond to the best and worst situations for the reception of PDM-SSB signal using a manually controlled PC, respectively. When $\alpha$ is 45°, the optical carrier is all located on the X-polarization while Y-polarization is carrier-less, which leads to the maximum polarization crosstalk. Under 12 dB CSPR, we measure the BERs as a function of launch power at different polarization angles, shown in Fig. 7. The optimal launch power is 7 dBm. There is a BER gap between $\alpha=0°$ and $\alpha=45°$, which results from electrical noise. This penalty from the interaction between polarization rotation and electrical noise also exists in SSFR [16], which can be mitigated using electrical amplifiers.

Using the optimal launch power of 7 dBm, we measure the BERs versus OSNR with different polarization angles at BTB and 80-km transmission case, shown in Fig. 8. At BTB case, the OSNR gap between SP and DP signal is only 3.5 dB, which is slightly greater than the 3 dB theoretical penalty. The results of BTB and 80-km transmission imply that the system is resistant to chromatic dispersion since the optical field is digitally reconstructed in the receiver DSP, and thus the CD can be compensated. For 80-km transmission, the OSNR penalty induced by the interaction between electrical noise and polarization rotation is about 1.1 dB and becomes smaller in the low-OSNR regime where the optical noise dominates the system performance.

The net bit rate is calculated as 382.8 ($= 56 \times 4 \times 2/1.14/(512 + 80 + 22320) \times 22320$) Gb/s with consideration of both frame redundancy and the 14% overhead hard-decision forward error correction (HD-FEC) [29]. Note that the CSPR can be further lowered if the edge-carrier assisted phase retrieval (ECA-PR) scheme [8] and phase retrieval receiver are adopted to substitute the KKR and SPDs here.

*C. Discussion*

Qualitatively, from the perspective of the modulation dimension, we compare various schemes of SSFR and JSFR in terms of hardware complexity. The results are summarized in Table I. For 2-D modulation, JSFR requires less hardware to recover the PDM-SSB signal than SSFR. One coupler and SPD can be saved at most. For 4-D JSFR, the hardware complexity depends on GR. We take the two-branch phase retrieval receiver [5-8, 23] for example, 6 or 8 SPDs are required to achieve optical full-field recovery. A simple method to achieve to 4-D JSFR (Scheme J-4D-1/2/3) using 3 GRs are analyzed in APPENDIX B, and will be investigated in our future work. For 4-D SSFR, 4 BPDs and 2 SPDs are required [22]. Therefore, the hardware complexity of SSFR and JSFR are almost the same for optical full-field recovery.

V. CONCLUSION

We propose a Jones-space field recovery (JSFR) scheme to address the polarization fading issue and achieve polarization diversity for single polarization carrier-assisted direct detection schemes. The different receiver structures of JSFR and the simplification method for JSFR are derived theoretically. Simulations are conducted to investigate the penalty from polarization fading, the CSPR condition for JSFR, and the impact of PMD on JSFR. In the concept-of-proof experiment, we demonstrate 56 GBd DP-16-QAM transmission over 80-km SMF using proposed 2-D JSFR and KKR-based single-polarization optical field recovery. Moreover, we make a qualitative comparison between Stokes-space field recovery and Jones-space field recovery in terms of the modulation dimension and hardware complexity. The proposed JSFR can realize dual-polarization optical field recovery and has potential to exploit higher dimensions, providing a low-cost solution for short-reach optical networks.

## APPENDIX

### A. JSFR based on optical hybrid and 3×3 coupler

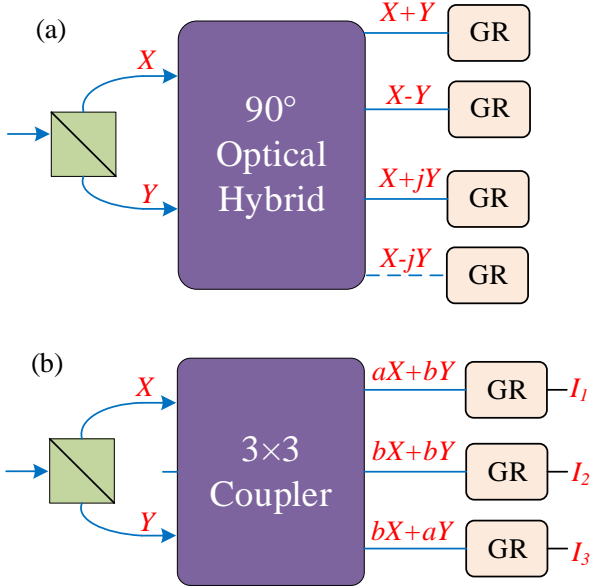

Fig. 9. Proposed JSFR based on (a) 90° optical hybrid and (b) 3×3 coupler.

In this subsection, we introduce the principle of JSFR based on optical hybrid and 3×3 coupler. The receiver structure based on optical hybrid is shown in Fig. 9(a). The branches $X$ and $Y$ are fed into the 90° optical hybrid, providing four outputs of $X+Y$, $X-Y$, $X+jY$, and $X-jY$, respectively. The four optical fields can be expressed as:

$$\begin{aligned}
X+Y &= 2\cos\alpha\cos\theta C + (\cos\alpha e^{j\theta}+\sin\alpha)S_x + (\cos\alpha e^{-j\theta}-\sin\alpha)S_y \\
X-Y &= (-2\sin\alpha+2\cos\alpha j\sin\theta)C + (\cos\alpha e^{j\theta}-\sin\alpha)S_x \\
&\quad -(\cos\alpha e^{-j\theta}+\sin\alpha)S_y \\
X+jY &= \{\cos\alpha\cos\theta-\sin\alpha+\cos\alpha\sin\theta \\
&\quad + j(\cos\alpha\cos\theta+\sin\alpha+\cos\alpha\sin\theta)\}C \\
&\quad +(\cos\alpha e^{j\theta}+j\sin\alpha)S_x + (j\cos\alpha e^{-j\theta}-\sin\alpha)S_y \\
X-jY &= \{\cos\alpha\cos\theta-\sin\alpha-\cos\alpha\sin\theta \\
&\quad + j(\cos\alpha\sin\theta-\sin\alpha-\cos\alpha\cos\theta)\}C \\
&\quad +(\cos\alpha e^{j\theta}-j\sin\alpha)S_x - (j\cos\alpha e^{-j\theta}+\sin\alpha)S_y
\end{aligned} \quad (10)$$

The CSPRs of the four optical fields can be calculated as:

$$\begin{aligned}
CSPR_{X+Y}(\alpha,\theta) &= 2\cos^2\alpha\cos^2\theta \cdot C_{req} \\
CSPR_{X-Y}(\alpha,\theta) &= (2-2\cos^2\alpha\cos^2\theta)\cdot C_{req} \\
CSPR_{X+jY}(\alpha,\theta) &= (1+2\cos^2\alpha\cos\theta\sin\theta)\cdot C_{req} \\
CSPR_{X-jY}(\alpha,\theta) &= (1-2\cos^2\alpha\cos\theta\sin\theta)\cdot C_{req}
\end{aligned} \quad (11)$$

Similarly, it is easy to prove that two of $CSPR_{X+Y}$, $CSPR_{X-Y}$, $CSPR_{X+jY}$ and $CSPR_{X-jY}$ are greater than $C_{req}$. Therefore, the full-field can be recovered using GRs and corresponding optical field recovery algorithms. The simplification method in Section II D can also be used in the JSFR based on 90° optical hybrid, given as:

$$\begin{aligned}
|(X-jY)\otimes h(t)|^2 &= 2(|X\otimes h(t)|^2+|Y\otimes h(t)|^2)-|(X+jY)\otimes h(t)|^2 \\
&= |(X+Y)\otimes h(t)|^2+|(X-Y)\otimes h(t)|^2-|(X+jY)\otimes h(t)|^2
\end{aligned} \quad (12)$$

Then we focus on the JSFR base on 3×3 coupler. The receiver structure is displayed in Fig. 9(b). After $X$ and $Y$ are mixed in the 3×3 coupler [30], the three outputs can be expressed as:

$$\begin{bmatrix} aX+bY \\ bX+bY \\ bX+aY \end{bmatrix} = \begin{bmatrix} a & b & b \\ b & a & b \\ b & b & a \end{bmatrix}\begin{bmatrix} X \\ 0 \\ Y \end{bmatrix}, \quad (13)$$

where $a$ and $b$ are $(2e^{j2\pi/9}+e^{-j4\pi/9})/3$ and $(e^{-j4\pi/9}-e^{j2\pi/9})/3$, respectively. The three corresponding photocurrents are denoted as:

$$I_1 = |aX+bY|^2, I_2 = |bX+bY|^2, I_3 = |bX+aY|^2 \quad (14)$$

we can digitally reconstruct four photocurrents ($|X+Y|^2$, $|X-Y|^2$, $|X+jY|^2$ and $|X-jY|^2$) using $I_1$, $I_2$, and $I_3$, which is shown as:

$$\begin{aligned}
|X+Y|^2 &= 3I_2 \\
|X-Y|^2 &= 2I_1 - I_2 + 2I_3 \\
|X+jY|^2 &= (1-\sqrt{3})I_1 + I_2 + (1+\sqrt{3})I_3 \\
|X-jY|^2 &= (1+\sqrt{3})I_1 + I_2 + (1-\sqrt{3})I_3
\end{aligned} \quad (15)$$

The reconstruction procedure can be achieved employing an analog circuit instead of DSP circuit, which avoids the electrical and quantization noise from DSO. The reconstructed photocurrents can be considered as the photocurrents from $X+Y$, $X-Y$, $X+jY$, and $X-jY$, and thus the full-field can be recovered using JSFR based on 3×3 coupler and 3 GRs.

### B. Simplification for JSFR by enhancing the carrier power

In this subsection, we introduce the simplification method for JSFR based on enhancing the carrier power. As derived above, the outputs of 3×3 coupler are $aX+bY$, $bX+bY$ and $bX+aY$, respectively. Thus, we calculate the CSPRs of these three optical fields ($aX+bY$, $bX+bY$ and $bX+aY$), given as:

$$\begin{aligned}
CSPR_{aX+bY}(\alpha,\theta) &= (1+3\cdot\mathrm{Re}\{ab^*(\cos^2\alpha e^{j2\theta}-\sin^2\alpha)\})\cdot C_{req}, \\
CSPR_{bX+bY}(\alpha,\theta) &= (2\cos^2\alpha\cos^2\theta)\cdot C_{req}, \\
CSPR_{bX+aY}(\alpha,\theta) &= (1+3\cdot\mathrm{Re}\{ba^*(\cos^2\alpha e^{j2\theta}-\sin^2\alpha)\})\cdot C_{req}
\end{aligned} \quad (16)$$

To guarantee the DP optical field after polarization rotation can be recovered by MIMO equalization in the receiver DSP, at least two of the three optical fields should satisfy the CSPR requirement for optical field recovery (CSPR>$C_{req}$). The second largest function is denoted as:

SecondMax($\alpha,\theta$)=
SORT$\left[\{CSPR_{aX+bY}(\alpha,\theta),CSPR_{bX+bY}(\alpha,\theta),CSPR_{bX+aY}(\alpha,\theta)\},2\right]$, (17)

where SORT is the sorting function. Thus, SecondMax($\alpha, \theta$) is also a function of $\alpha$ and $\theta$. It is easy to find that the minimum of SecondMax($\alpha, \theta$) is 0.5·$C_{req}$. Therefore, the DP optical full-field can be recovered using the proposed JSFR and corresponding GR given doubled carrier power, or 3 dB higher CSPR equivalently. We take an arbitrary SOP for an example. When $\alpha$ and $\theta$ are both $\pi/16$, $CSPR_{aX+bY}$, $CSPR_{bX+bY}$ and $CSPR_{bX+aY}$ are 0.51·$C_{req}$, 3.7·$C_{req}$, and 1.88·$C_{req}$, respectively. Thus, the optical fields ($bX+bY$, $bX+aY$) can be recovered, and then the DP optical full-field can be recovered using MIMO equalization. Therefore, by enhancing the initial CSPR to 2·$C_{req}$ (3 dB), the analog circuit can be further eliminated in JSFR based on 3×3 coupler.

Similarly, this simplification method can be used to JSFR based on 2×2 coupler and 90° optical hybrid. However, the minimum of SecondMax($\alpha, \theta$) is (1-$\sqrt{2}$/2)·$C_{req}$. The initial CSPR should be enhanced to (2+$\sqrt{2}$)·$C_{req}$ (~5.3 dB). So, the simplification for JSFR based on 3×3 coupler is more recommended since only 2·$C_{req}$ is required. In summary, only 3 GRs are required for JSFR by enhancing the initial CSPR to (2+$\sqrt{2}$)·$C_{req}$ or 2·$C_{req}$.